\documentclass[10pt]{iopart}
\usepackage{graphicx}
\usepackage{xspace}				
\usepackage[pdftex]{color}			

\newcommand{\bma}[1]{\mbox{\boldmath$#1$}}

\newcommand{\dmu}{\textsl{DMU}\xspace}
\newcommand{\dds}{\textsl{DDS}\xspace}
\newcommand{\esa}{\textsl{ESA}\xspace}
\newcommand{\ieec}{\textsl{IEEC}\xspace}
\newcommand{\ifae}{\textsl{IFAE}\xspace}
\newcommand{\lca}{\textsl{LCA}\xspace}
\newcommand{\lisa}{\textsl{LISA}\xspace}
\newcommand{\lpf}{\textsl{\mbox{LPF}}\xspace}
\newcommand{\ltp}{\textsl{LTP}\xspace}
\newcommand{\nasa}{\textsl{NASA}\xspace}
\newcommand{\nte}{\textsl{NTE}\xspace}

\begin{document}


\jl{6}

\title[The diagnostics subsystem on board \textsl{LISA PathFinder}\ldots]{The
diagnostics subsystem on board \textsl{LISA PathFinder} and \textsl{LISA}}

\author{P~Ca\~nizares$^{1,4}$, A~Conchillo$^{1,4}$, E~Garc\'\i a-Berro$^{2,4}$,
L~Gesa$^{1,4}$, C~Grimani$^3$, I~Lloro$^{1,4}$, A~Lobo$^{1,4}$,
I~Mateos$^{1,4}$, M~Nofrarias$^5$, J~Ramos-Castro$^6$, J~Sanju\'an$^{1,4}$
and Carlos~F Sopuerta$^{1,4}$}

\address{$^1$ Institut de Ci\`encies de l'Espai, {\sl CSIC}, Facultat de
 Ci\`encies, Torre C5 parell, 08193 Bellaterra, Spain}

\address{$^2$ Departament de F\'\i sica Aplicada, UPC, Escola Polit\`ecnica
 Superior de Castelldefels,  Avda.\ del Canal Ol\'\i mpic s/n,
 08860 Castelldefels, Spain}

\address{$^3$ Universit\`a degli Studi di Urbino, and {\sl INFN\/} Florence,
Istituto di Fisica, Via Santa Chiara 27, 61029 Urbino, Italy}

\address{$^4$ Institut d'Estudis Espacials de Catalunya (IEEC), Edifici Nexus,
Gran Capit\`a 2-4, 08034 Barcelona, Spain}

\address{$^5$Max-Planck-Institut f\"ur Gravitationsphysik
 (Albert-Einstein-Institut), Callinstrasse 38, D-30167 Hannover, Germany}

\address{$^6$Departament d'Enginyeria Electr\`onica, UPC, Campus Nord,
 Edifici C4, Jordi Girona 1-3, 08034 Barcelona, Spain}

\ead{lobo@ieec.fcr.es}

\date{11-August-2008}

\begin{abstract}
The Data and Diagnostics Subsystem of the \ltp hardware and software are at
present essentially ready for delivery. In this presentation we intend to
describe the scientific and technical aspects of this subsystem, which
includes thermal diagnostics, magnetic diagnostics and a Radiation Monitor,
as well as the prospects for their integration within the rest of the \ltp.
We will also sketch a few lines of progress recently open towards the more
demanding diagnostics requirements which will be needed for \lisa.
\end{abstract}
\submitto{\CQG}
%

\section{Introduction}
\label{lobo-sec1}

\lisa is a technologically sophisticated mission. In its current
baseline design, an arm length of 5 million kilometers is envisaged,
and its acceleration noise is required to satisfy~\cite{lobo-lisasrd}
%
%
\begin{equation}
 \hspace*{-6em}
 S_{\delta a, {\rm LISA}}^{1/2}(\omega)\leq 3\times 10^{-15}\,\left\{\left[
 1 + \left(\frac{\omega/2\pi}{8\ {\rm mHz}}\right)^{\!\!4}\right]\!
 \left[1 + \left(\frac{0.1\ {\rm mHz}}{\omega/2\pi}\right)\right]
 \right\}^{\!\frac{1}{2}} \ {\rm m}{\rm s}^{-2}/\sqrt{\rm Hz}
 \label{lobo-eq1}
\end{equation}
in the frequency band between 0.1 mHz and 0.1 Hz.

\textsl{LISA PathFinder} (\lpf) has a reduced acceleration noise budget,
both in magnitude and in frequency band~\cite{lobo-lpfsrd},
\begin{equation}
 S_{\delta a, {\rm LPF}}^{1/2}(\omega)\leq 3\!\times\!10^{-14}\,\left[
 1 + \left(\frac{\omega/2\pi}{3\ {\rm mHz}}\right)^{\!\!2}\right]\,
 {\rm m}{\rm s}^{-2}/\sqrt{\rm Hz}
 \label{lobo-eq2}
\end{equation}
in the frequency band between 1 mHz and 30 mHz. This noise is the
result of various disturbances which limit the performance of the
instrumentation on-board. A number of these can be specifically
monitored and dealt with by means of suitable devices, which form
the so called \emph{Diagnostics Subsystem}. In the case of \lpf,
these include thermal and magnetic diagnostics, plus the Radiation
Monitor (RM), which provides counting and spectral information on
ionising particles hitting the proof masses. The purpose of this
note is to summarise the latest developments on the \lpf Diagnostics
Subsystem, developed in Barcelona, including some preliminary
research results on the extension of their performance in view of
the future \lisa. The diagnostics in \lpf, as a technology precursor
of \lisa, are intended to help design a quieter environment in the
\lisa spacecraft. Their role in \lisa is still to be defined, but
they will likely work as a noise debugging tool, much in the same
spirit as in \lpf, which will provide house-keeping data and assist
in GW signal dig-out.

Background to the diagnostics motivation and main requirements will
be omitted here, but the reader will find details in~\cite{lobo-ere2006}
and references therein. In this paper we will sequentially review
the latest relevant results on each of the diagnostics items.

\section{Thermal diagnostics}
\label{lobo-sec2}

\begin{figure}
\centering
\includegraphics[width=0.49\columnwidth]{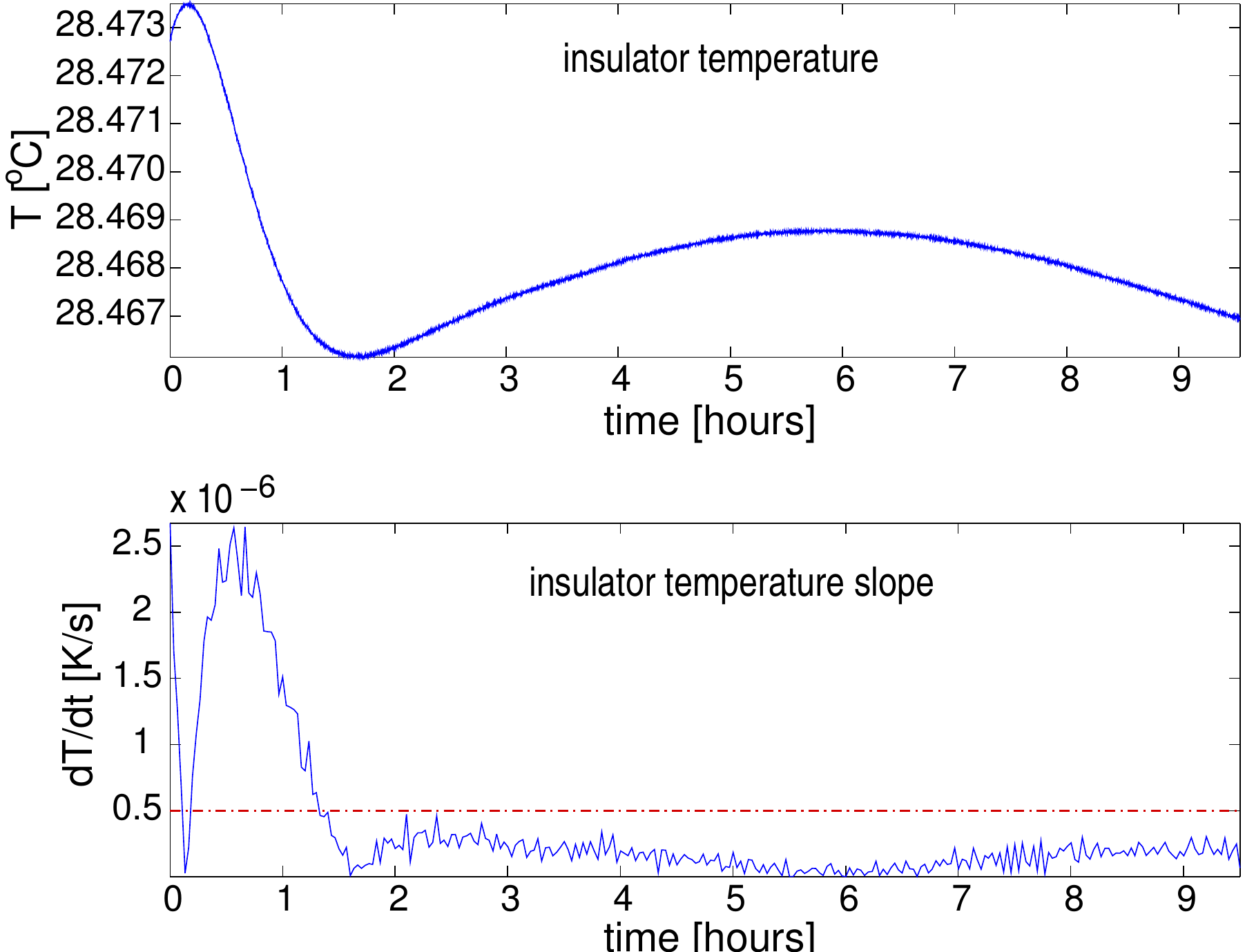}\ \ 
\includegraphics[width=0.49\columnwidth]{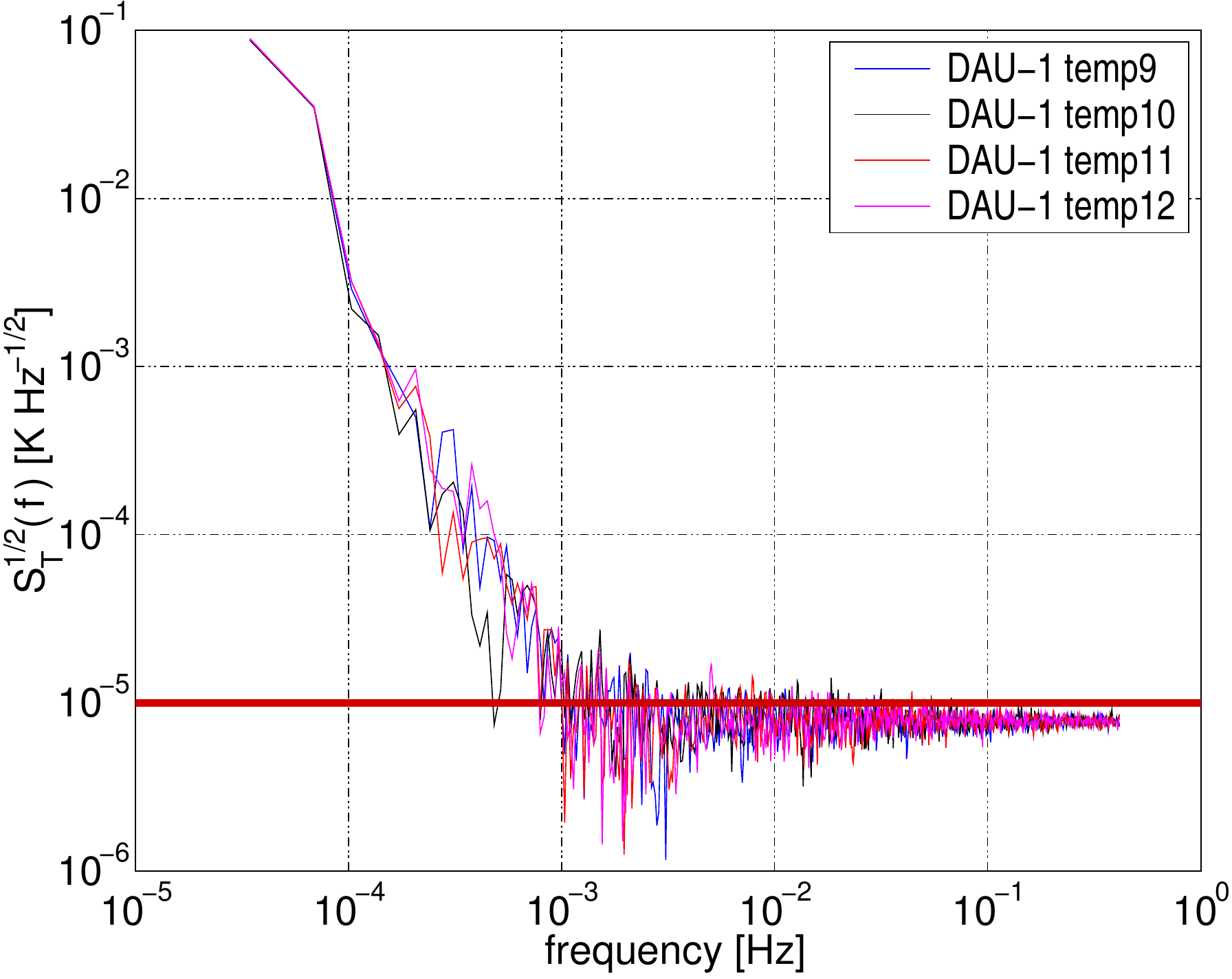}
\caption{Flight model temperature sensor behaviour. Left panel: 9.5
 hour data run, as read by the NTCs. The upper graph shows the raw
 data, and the lower graph its time variations. Note the slope is
 below the critical value of 0.5\,$\mu$K/sec only when the testbed
 has reached thermal stability after almost two hours. Right panel:
 spectral density of noise for 4 NTCs: all of them are below the
 required 10$^{-5}$\,K\,Hz$^{-1/2}$ threshold throughout the
 measuring bandwidth. Data to produce the plot correspond to the
 period of low temperature slope.
\label{lobo-fig1}}
\end{figure}

The temperature stability required to prevail inside the \lca (\ltp Core
Assembly, \ltp = \lisa Technology Package) is, in spectral density of
temperature fluctuations, 10$^{-4}$\,K\,Hz$^{-1/2}$ in the measuring
bandwidth. Studies carried through at \ieec during the prototyping stage
determined that the only option compatible with reliable temperature
measurements at that level was the use of thermistor devices ---or NTC,
Negative Temperature Coefficient devices~\cite{lobo-fee}. After successful
verification that the NTCs plus their front-end electronics worked OK, the
circuitry was integrated in the \dmu (Data Management Unit, the \ltp computer),
and flight hardware and software were recently submitted to further
test. The results are shown in Figure~\ref{lobo-fig1}.

During the data taking run, the sensors were placed inside an insulator
jig which strongly damps any ambient temperature fluctuations. The jig
consists in an aluminium metal core, where the NTCs are attached,
surrounded by a thick layer of polyurethane, with a very low thermal
conductivity coefficient~\cite{lobo-pep}. The damping efficiency of
the device is large enough for thermal screening of its interior,
but the connecting harness between the sensors and the electronics
outside constitutes a leakage line which does degrade in practice
the conditions for the test. In order to maintain the temperature
of the NTCs stable over long periods of time, a temperature feedback
control was added to the system ---see next paragraph. Another factor
of improvement was to keep ambient temperature as stable as feasible.
For this, the experiment was done inside a well insulated anechoic
chamber in the Institute's building basement floor. To ensure
temperature stability conditions, the whole setup (insulating jig,
front-end electronics and computers) was locked and left untouched
for two days before starting the experiment.

The \dmu has two identical redundant DAUs (Data Acquisition Units), and
the plots in~\ref{lobo-fig1} reflect the data taken by DAU-1. A very
important circumstance has to be taken into consideration when the data
analysis is performed. This is the temperature drifts, which need to be
maintained small, more specifically, $|dT/dt|<$0.3\,$\mu$K/sec. The
reason is the non-linear behaviour of the ADC (Analog-to-Digital
Converter), which introduces spurious noise at low frequencies due
to quantisation errors. The ADC has 16 bits, and the effect could
be avoided with a larger bit depth ADC ---which is not available for
flight. The temperature feedback control system in the jig's aluminium
core mentioned above takes care of the stability conditions of the NTCs
temperature at very low frequencies~\cite{lobo-thtp}, thereby enlarging
the length of measuring time available for analysis. Usable data in the
reported run were 8 hours (see left panel, lower plot), not a very long
stretch yet widely sufficient to obtain a reliable spectral estimation
down to 1\,mHz.

The spectra shown in the right panel correspond to four sensors (labelled
9 through 12), attached to one of the three multiplexer boards in the DAU.
In order to detect temperature fluctuations in the \lca below the stability
conditions of 10$^{-4}$\,K\,Hz$^{-1/2}$, a requirement was set on the
temperature sensing of 10$^{-5}$\,K\,Hz$^{-1/2}$~\cite{lobo_ddssr}. As
can be seen, all these sensors perform according to the requirement in
the entire \ltp band, from 1\,mHz to 30\,mHz.

\subsection{Looking into \textsl{LISA}}
\label{lobo-sec2.1}

Noise steeply rises towards lower frequencies, which in turn poses the
question of how difficult it may be to reach the lower frequency band
of \lisa with suitable sensitivity at 0.1\,mHz. Recent research work
at \ieec has shown that neither the current electronics design nor the
sensors themselves are limiting factors. Rather, it is the experimental
conditions, described above, which appear to be unsuitable to properly
assess the real performance capabilities of the thermal sensing system:
heat leakage through the wiring and insufficient screening capacities of
the surrounding jig have proved to be at the root of the problem, instead.
Preliminary experiments with an improved wiring concept, and use of
\emph{differential} temperature measurements has shown that it is already
possible to reach a level of noise below 10$^{-5}$\,K\,Hz$^{-1/2}$ at
0.1\,mHz with no changes to the electronics and with the same NTCs. It is
conjectured that even 1\,$\mu$K\,Hz$^{-1/2}$ can be attainable. Although
\lisa requirements are still not fully defined in this area, these results
are really promising. The reader is referred to Sanju\'an's contribution
to this volume for further details and plots on this important matter.

\subsection{Heaters}
\label{lobo-sec2.2}

The availability of excellent thermometers is not very useful
of itself. Actually, their use is to convert temperature fluctuation
information into test mass acceleration noise. In other words, we would
like to know which fraction of the total \ltp readout noise is due to
temperature fluctuations. For this, calibration is needed, which will
translates temperature measurements to acceleration readout. In the \ltp,
the procedure to obtain such relationship is the use of controlled, high
signal-to-noise ratio thermal signals applied in suitably chosen
locations, and measure the observed system response in parallel with
temperature measurements. A (matrix) \emph{transfer function} is thus
obtained, which can subsequently be applied to the thermometers' readings
to determine the specific weight of temperature noise in the \ltp total
noise~\cite{lobo-ere2006}.

Several modelling and laboratory experiments have been done to characterise
the above procedure, with very interesting results~\cite{lobo-ow,lobo-mnt}.
In flight, the analysis is more complicated, as the calibration process
interacts with the full \ltp dynamics loop. The on-ground experiment
analysis results can be directly fed into that loop, and the mission
master plan naturally includes suitable protocols to deal with the
heater signals and the inference of the corresponding transfer functions.
The reader is referred to Nofrarias's contribution to the JPCS volume of
this Proceedings for the latest progress on this matter.

\section{Magnetic diagnostics}
\label{lobo-sec3}

Again, this section only reports on the latest results on \ltp
magnetic diagnostics. The reader will find background information
in~\cite{lobo-ere2006}. The \ltp Test Masses (TMs) are two cubes 4.6~cm
to the side, weighing 1.96~kg each. They are made of an alloy of gold
and platinum with 70\,\% Au + 30\,\%\ Pt. To cast such an alloy is a
process where ferromagnetic impurities can contaminate the alloy
structure, thus leaving a remanent magnetic moment ${\bf m}_0$ in the
TM. Likewise, magnetic susceptibility, $\chi$,  will be present. These
are required to comply with the following constraints~\cite{lobo-lpfsrd}:
\begin{equation}
 |\chi| < 10^{-5}\ ,\quad
 |{\bf m}_0| < 10^{-8}\ {\rm Am}^2
 \label{lobo-eq3}
\end{equation}

\begin{figure}
\centering
\includegraphics[width=0.5\columnwidth]{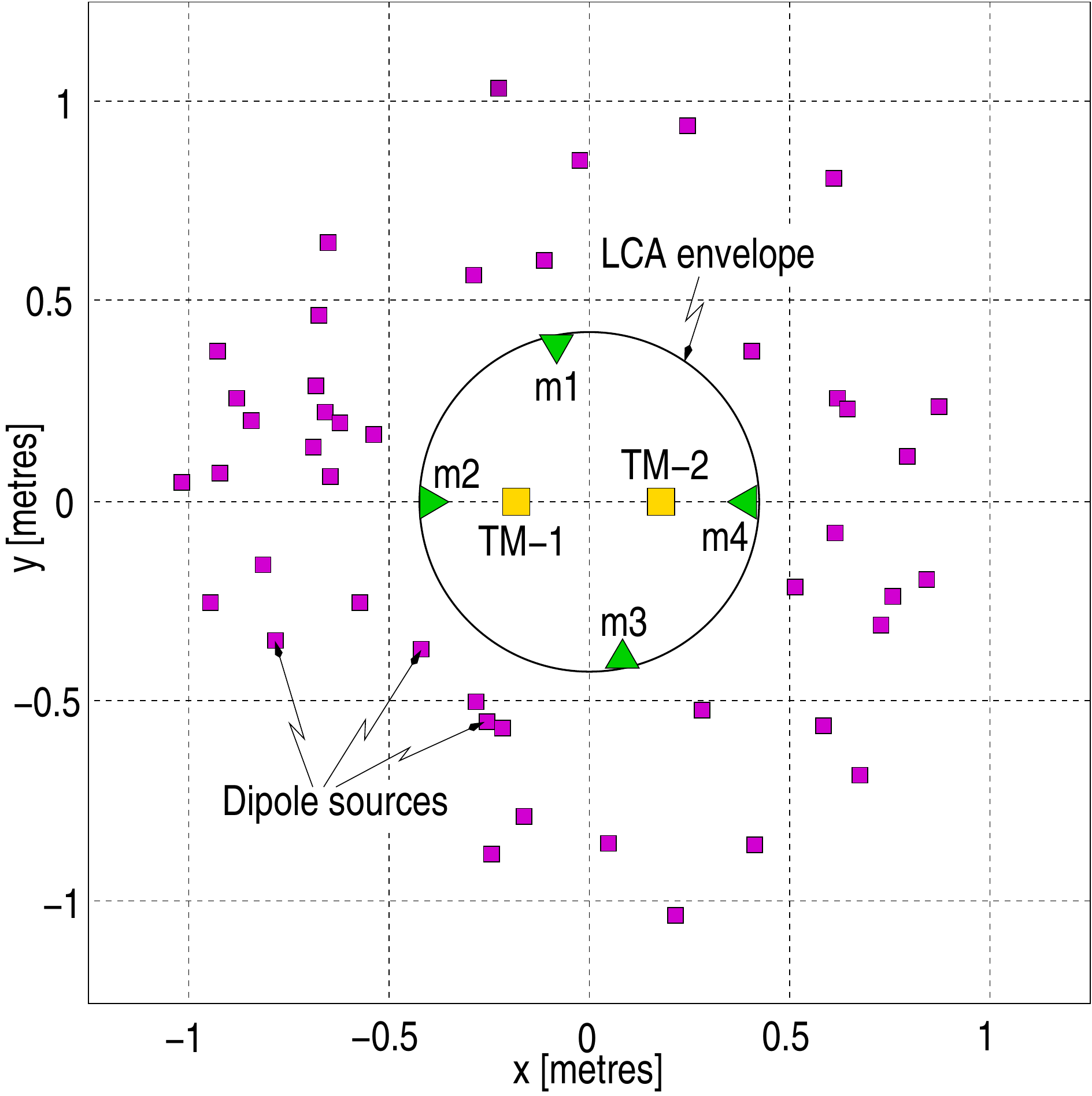}
\caption{Projection view of the dipole magnetic sources in the \lpf
 spacecraft (small squares), the TMs (larger squares) and the
 magnetometers (triangles).
\label{lobo-fig2}}
\end{figure}

In spite of these low values, magnetic field and gradient fluctuations
in the TMs result in spurious forces on them, causing acceleration noise
which adds undiscriminated to the \ltp readout. In order to diagnose the
state of the magnetic environment, a set of high sensitivity vector
magnetometers are placed in the \lca wall ---see figure~\ref{lobo-fig2}.

These magnetometers are tri-axial \emph{fluxgate} magnetometers which
have a relatively large \emph{Permalloy} core. They are very sensitive
---see below---, but should be kept somewhat far from the TMs to avoid
magnetic back action disturbances on the latter.

Figure~\ref{lobo-fig2} also displays the positions of identified sources
of magnetic field in the \lca~\cite{lobo-wealthy}. These sources are all
\emph{beyond} the \lca walls, and come from various circuitry and other
magnetic components in the spacecraft. Clearly, the magnetic field in the
TMs is smaller than it is in the magnetometers, since it decays towards
the inner region of the \lca as it is produced by magnetic dipoles. This
poses a problem of interpolation between the magnetometers' readouts to
obtain the actual field in the TMs positions. We now describe how to
address this problem, according to our current understanding.

\subsection{Magnetic field interpolation}
\label{lobo-sec3.1}

We assume the TMs are small size compared to the \lca volume, hence we
consider the magnetic field, ${\bf B}({\bf x})$ inside the \lca to be
mostly a \emph{vacuum field}. This means
$\bma{\nabla}$$\times$${\bf B}$\,=\,$\bma{\nabla\cdot}{\bf B}$\,=\,0, or
\begin{equation}
 {\bf B}({\bf x}) = -\bma{\nabla}\Psi({\bf x})\ ,\quad
 \nabla^2\Psi({\bf x}) = 0
 \label{lobo-eq4}
\end{equation}
where $\Psi({\bf x})$ is a scalar function. If this is expanded in terms
of spherical harmonics, $Y_{lm}$, then ${\bf B}({\bf x})$ ensues as
%
%
%
\begin{equation}
 {\bf B}({\bf x}) = \frac{\mu_0}{4\pi}\,\sum_{l=1}^{\infty}\,\sum_{m=-l}^{l}\,
 M_{lm}\,\bma{\nabla}\left[r^lY_{lm}({\bf n})\right]\ ,\quad
 r\equiv |{\bf x}|\ ,\quad {\bf n}\equiv {\bf x}/r
 \label{lobo-eq6}
\end{equation}
where $M_{lm}$ are the \emph{multipole coefficients} of the expansion.
Ideally, an infinite number of coefficients are necessary to calculate
${\bf B}({\bf x})$, which would be possible if the field was known in
all points of a closed surface not containing any field source\footnote{
Strictly speaking, this is not correct: indeed, having as many sensors
as dipoles there are would suffice, as we know the magnetic field is
generated by a finite set of such dipoles.}.
In real practice, we only have four magnetometers, so the number of
multipole coefficients we can actually determine is fixed by this
circumstance. The counting is easy to~do: there are 12 sensor data
channels, three per magnetometer (recall they are tri-axial). The
number of $M_{lm}$ we can calculate is accordingly 12, or less.
There is no monopole contribution to {\bf B}, of course, there are
three dipole coefficients, five quadrupole, seven octupole, etc. The
series expansion in equation~\eref{lobo-eq6} must thus be cut at
$l$\,=\,2, since continuing it up to $l$\,=\,3 would require
3\,$+$\,5\,$+$\,7\,=\,15 $M_{lm}$, but we can only estimate~12.
The following is thus our best approximation:
\begin{equation}
 {\bf B}_{\rm approx}({\bf x}) = \frac{\mu_0}{4\pi}\,
 \sum_{l=1}^{2}\,\sum_{m=-l}^{l}\,M_{lm}\,
 \bma{\nabla}\left[r^lY_{lm}({\bf n})\right]
 \label{lobo-eq7}
\end{equation}

Next task is to determine the 8 dipole $+$ quadrupole coefficients. This
we do by a least square method, where we define the square error as
\begin{equation}
 \varepsilon^2(M_{lm}) = \frac{1}{2}\,\sum_{s=1}^4\,
 \left|{\bf B}_{\rm approx}({\bf x}_s) -
       {\bf B}_{\rm measured}({\bf x}_s)\right|^2
 \label{lobo-eq8}
\end{equation}
where $s\/$ is an index labelling the magnetometers, located at positions
${\bf x}_s$, and ${\bf B}_{\rm measured}({\bf x}_s)$ is the (vector)
readout of the $s\/$-th magnetometer. By solving the system of equations
$\partial\varepsilon^2/\partial M_{lm}$\,=\,0 we find $M_{lm}$ for
$l\/$\,=\,1,2, $m\/$\,=\,$-l,\ldots,l$. Feeding them into
equation~\eref{lobo-eq7} with {\bf x}\,=\,${\bf x}_{\rm TM}$, the positions
of the test masses, we get the desired interpolation to estimate the field
values at the TMs.

In order to verify the efficiency of this analysis, the following procedure
was implemented: series of magnetic moments of the dipole sources were
simulated randomly ---with some constraints on their moduli, as indicated
by the estimates at Astrium-Stevenage~\cite{lobo-wealthy}---, and the field
reconstruction algorithm was subsequently applied to each series. Then the
values of the so reconstructed field at the TMs were compared with the
exact ones, also case by case. Unfortunately, the results appear to be
poor: deviations between obtained and expected figures vary between
quite good (less than~10\%) and rather disappointing (factors of 5 and
eventually more). The reason for this poor result is easy to discover:
the series~\eref{lobo-eq7} only provides a \emph{linear} interpolation
algorithm between field values at the \lca boundary, where the sensors
are, to its interior, which cannot accurately account for the fact that
the field components have a trough somewhere there ---its position and
depth depending on the particular dipole distribution outside.

\subsection{New magnetic diagnostics concepts for \textsl{LISA}}
\label{lobo-sec3.2}

\begin{figure}
\centering
\includegraphics[width=0.49\columnwidth]{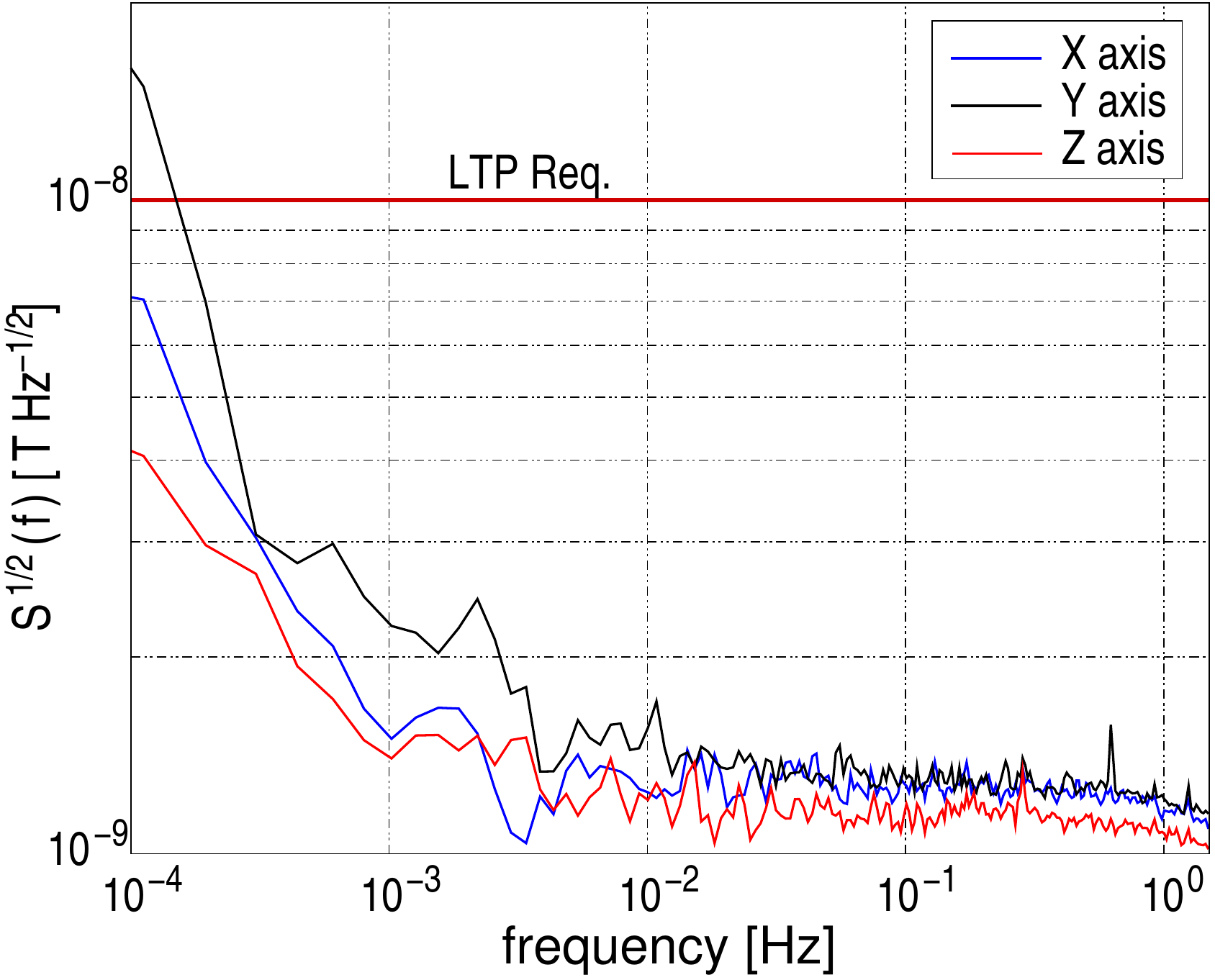}\ \ 
\includegraphics[width=0.49\columnwidth]{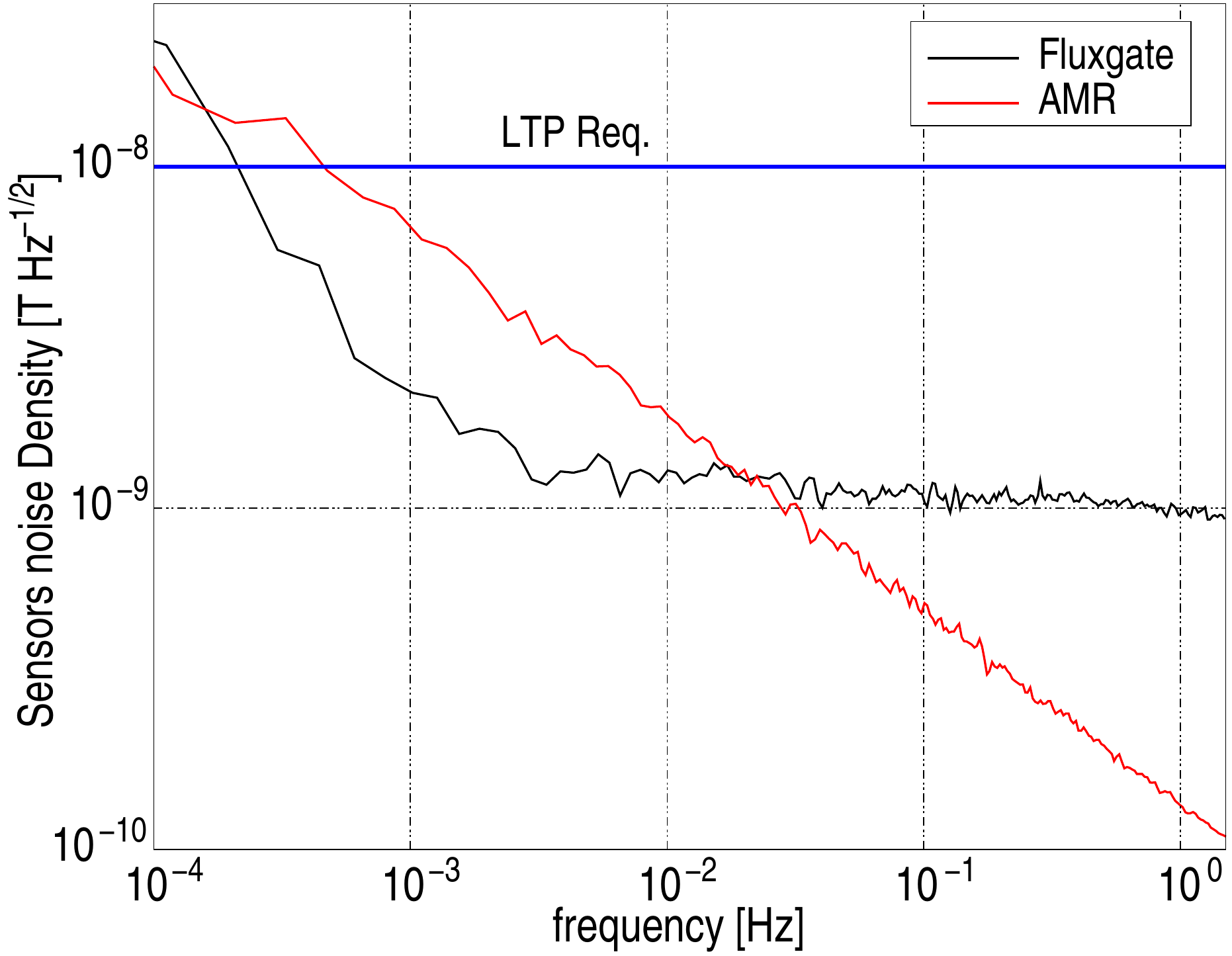}
\caption{Left panel: spectral density of noise of each of the three
channels of an \ltp fluxgate magnetometer. Right panel: comparison
of the performance of the latter with that an AMR magnetometer.
\label{lobo-fig3}}
\end{figure}

The magnetometers are required to have a level of noise below
10~nT\,Hz$^{-1/2}$ in the \ltp bandwidth. The fluxgates to be
flown on-board \lpf are comfortably compliant with that, as
shown in figure~\ref{lobo-fig3}, left panel.

Such excellent performance, however, does not quite match the quality
of the results we can derive from their output, as discussed in the
previous section. A more faithful reconstruction of the magnetic field
at the TM positions requires the magnetometers to be closer to them,
but fluxgates cannot be mounted there: back-action would be unacceptably
high, and space resolution is poor (the sensor heads are $\sim$2\,cm long).
An investigation of alternative solutions has just begun at \ieec to
improve magnetic diagnostics for \lisa, whereby AMR (Anisotropic
Magneto-Resistance) devices are being considered. These are very tiny,
and at least three or more per TM could be attached to the spacecraft
structure appreciably closer to the TMs without risk of back action
effects. Preliminary results on the performance of AMRs is shown in
figure~\ref{lobo-fig3}, right panel, which do look encouraging. Details
on this matter will be found in Mateos's contribution to the JPCS volume
of this Proceedings, where various aspects of the problem are addressed,
including magnetic properties of the AMRs.

\subsection{Control coils}
\label{lobo-sec3.3}

Like with thermometers, magnetic field measurements are of themselves of
little use. We need to convert magnetic field and gradient fluctuations
into \ltp acceleration noise. For this, controlled magnetic forces are
applied to the TMs by means of non-homogeneous fields generated by coils,
which will serve calibration purposes of the magnetometers' readouts
---see background information details in~\cite{lobo-ere2006}. Because of
the magnetic susceptibility of the TMs, magnetic forces on them depend on
coupling of the magnetic field to its gradient. Acceleration fluctuations
accordingly depend on magnetic field gradient fluctuations as well as on
field DC values. This in turn dictates that DC fields generated by the
coils should be very stable not to degrade their performance. Tests to
check the current stability requirements in the coils are underway at
the time of writing. Provisional results look so far satisfactory. Test
Reports will be formally written after full analysis of the data is complete.

\section{The Radiation Monitor}
\label{lobo-sec4}

\lpf will be stationed in a Lissajous orbit around Lagrange point L1,
some 1.5 million km away from the Earth in direction to the Sun. There,
the spacecraft will be exposed to various ionising radiations coming
from the Galaxy and from the Sun. Some of these charged particles, will
be stopped by the spacecraft structure surrounding the TMs, while others
will make it to the TMs. The latter are particles having energies above
a threshold of about 100~MeV/nucleon, as shown by detailed simulations
done at Imperial College~\cite{lobo-lpfsimul}. The excess charge deposited
in the TMs depends on the primary energy of the incoming particle, since
secondary particles are generated inside the TMs as the primary travels
across the TM volume. The charge deposit is of course a random process
which results in acceleration noise due to interactions with the electric
system which monitors the position of the TMs in their enclosure, to
fluctuations of the position of the TMs relative to the electrostatic
centre of the electrode housing, and to Lorentz interaction with the
environmental and interplanetary magnetic field~\cite{lobo-Henrique}.

The \ltp is equipped with a system of ultraviolet lamps which are needed
to purge (by photo-electric effect) the charge accumulated in the TMs.
By accurately matching the discharge rates to the charging rates the
noise due to charging can be minimised. However, the GRS (Gravitational
Reference Sensor) can only track charging rates by averaging over certain
periods of time. The Radiation Monitor (RM) is capable of measuring these
charging rates over significantly shorter periods ---see below---, thereby
producing data which will be used to help match the measured charging
rates to the discharging rates, or else to clean \ltp data by off-line
analysis~\cite{lobo-diana}. Further details on this will be found in
Diana Shaul's contribution to this Proceedings.

\begin{figure}
\centering
\includegraphics[width=0.34\columnwidth]{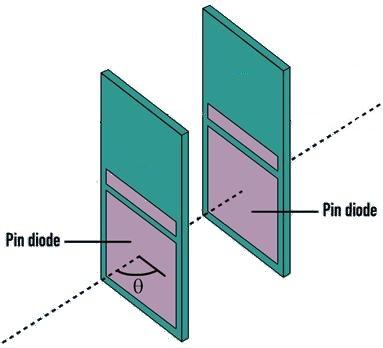}\ \ \ 
\includegraphics[width=0.56\columnwidth]{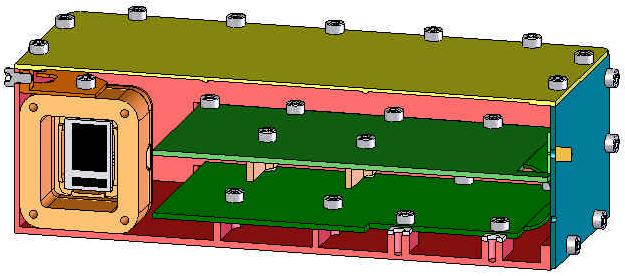}
\caption{Left panel: RM concept: two PIN-diodes in telescopic configuration
can both count single particles and measure their energy when events are
detected in coincidence. Right panel: actual hardware schematics. The
horizontal plates inside the box indicate the two PCBs, each connected
to one of the PINs.
\label{lobo-fig4}}
\end{figure}

The RM design was based on detailed simulations of the \lpf spacecraft
and the \ltp structure~\cite{lobo-lpfsimul}, and with a philosophy of
making it simple and light while, at the same time, being able to provide
not only particle counting but also spectral estimation to distinguish
Galactic Cosmic Rays (CGR) from Solar Energetic Particles (SEP). The
concept to implement such measurements is shown in figure~\ref{lobo-fig4},
left. Two silicon PIN diodes are placed parallel to each other in a
telescopic configuration\footnote{
The PIN diodes in the \ltp RM are spare samples from the Calorimeter PIN
Photodiode Assembly on board the \textsl{GLAST} mission, very kindly
supplied by Neil Johnson at no cost for us.}.
Each diode can count single particle events, but cannot tell whether the
particles were charged or not. Events detected in coincidence in both PINs
do instead correspond to charged particles, and their primary energy is
inferred form the energy deposition. There is however some uncertainty
here, due to degeneracy associated with the RM \emph{acceptance angle}:
higher energy particles with oblique incidence may deposit the same
energy as lower energy particles which impact perpendicular to the
PINs~\cite{lobo-cesar}.

\subsection{Technical details of the RM}
\label{lobo-sec4.1}

\begin{figure}
\centering
\includegraphics[width=0.56\columnwidth]{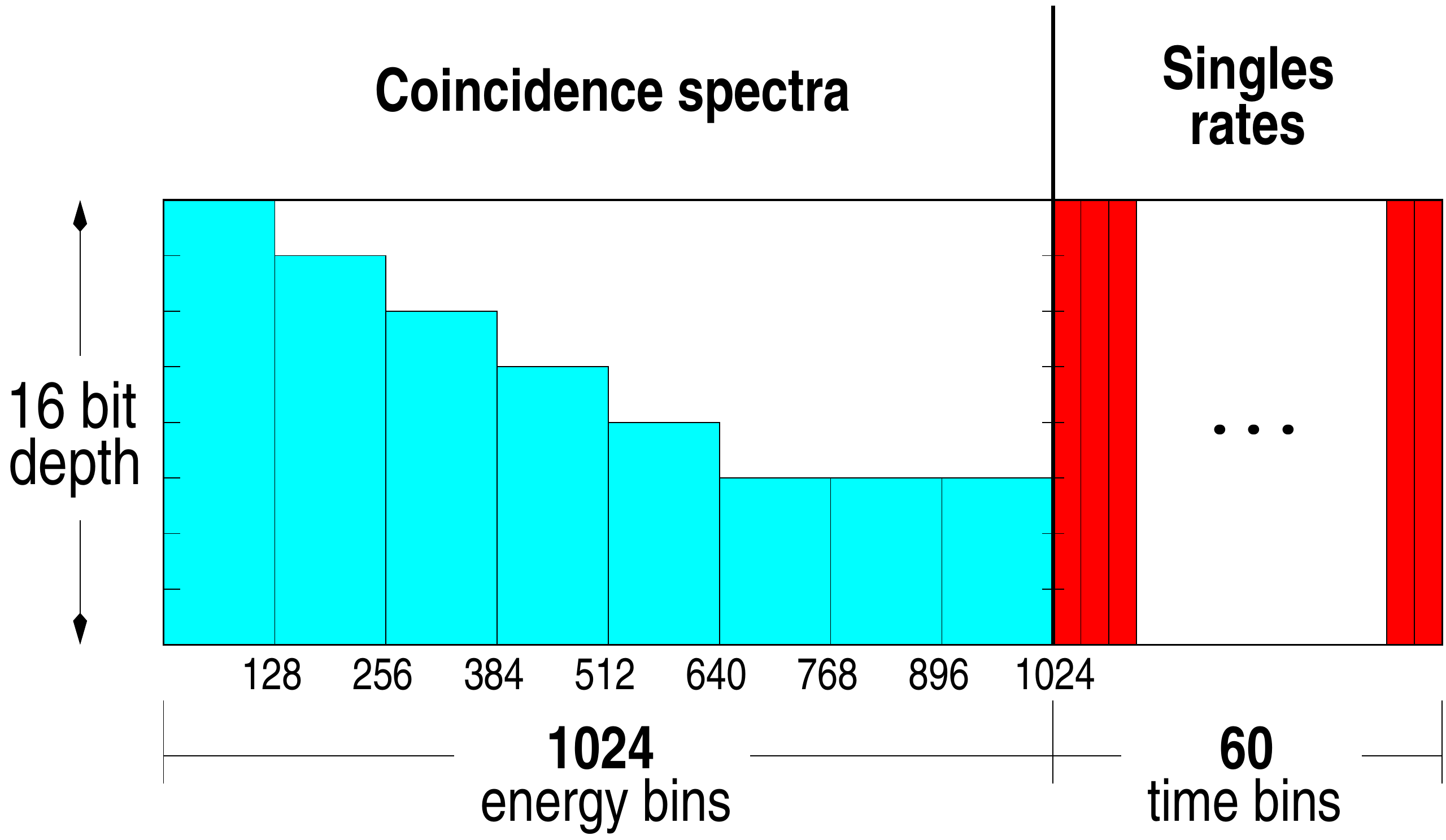}
\caption{RM data histogramme: the first 1024 bins contain deposited  energy
of coincident events, while the remaining 60 bins contain number of singles
counts, registered every 10.24 seconds. It takes 614.4 seconds to build up
a complete histogramme. Total bit rate is 17.8 bits/sec.
\label{lobo-fig5}}
\end{figure}

The RM delivers data accumulated over periods of 614.4 seconds
($\sim$\,10 minutes), and sends them to the \dmu in the form of
histogrammes~\cite{lobo-mokhtar}. Figure~\ref{lobo-fig5} displays the
structure of one of such histogrammes, with maximum bit depth in each bin,
according to the maximum foreseen event rates, both in single events and
coincidences, as well as in energy depositions. The latter range from~0
(actually $\sim$\,20 keV, due to noise) for the fastest incoming particles
to 5\,MeV for the slowest ones. Binning this range in 1024 equal length
intervals provides an energy resolution of nearly 5\,keV.

The RM ADC has 16 bits, and sampling rate is 100\,Hz. Data are accumulated
in memory until they are sent to the \dmu after 60 passages. In the first
passage, singles are counted and stored in the first singles bin, in the
second passage, singles are counted and stored in the second singles bin,
and so on. In each passage, the energy deposited by events detected in
coincidence in both PIN diodes is determined, the 1024 energy deposition
bins scanned, and the number of events stored in the corresponding energy
bins. Each passage therefore takes 10.24 seconds, hence 614.4 seconds are
needed to fill up the 60 passage histogramme.

The estimated average of RM GCR singles counts is 4 c/s, and 0.4 c/s for
coincident events. The largest SEP events observed so far can generate
up to a few thousand singles c/s, and about 10 times less in coincidence.
Therefore the RM electronics should be able to cope with such large
events without degrading its performance. A conventional 5000 c/s was
thus set as the requirement, which is comfortably satisfied by the RM:
indeed, each singles bin can accept up to 2$^{16}$\,=\,65,536 events
in 10.24 seconds, i.e., 6400 c/s. The depth of the energy bins seen in
figure~\ref{lobo-fig5} is uneven, based on the fact that large energy
depositions are less frequent than small ones. This was done to reduce
RM telemetry usage, but towards the end of May-2008 the \lpf Science
Working Team adopted a simplified scheme whereby all energy bins have
maximum depth, i.e., 16 bits, which does not entail any significant
increase in the mission telemetry budget.

The RM prototype underwent tests at the Proton Irradiation Facility
of the Paul Scherrer Institute (PSI) in late 2005 which proved fully
successful~\cite{lobo-peter}. Since then, the initial IFAE design was
used by \nte, the Spanish industrial contractor, to manufacture a RM EQM,
with a number of modifications to make it handy for integration and flight.
The EQM RM was recently debugged and green light for production of the FM
(Flight Model) was given. The FM is expected to be finished by the end of
2008, and it will be submitted to further tests at PSI to verify basic
functionality issues. The test will be milder than that done with the
prototype, as strong irradiation may damage the device.

\section{Conclusion}
\label{lobo-sec5}

The diagnostics subsystem of the \ltp will provide very useful noise
debugging information, which will help us understand the nature of
that noise, thence eventually guiding in various ways the progress
towards the improved sensitivity needed for \lisa.

The \ltp diagnostics subsysten must comply with a number of requirements
on sensitivity and performance, which have been implemented and tested to
satisfaction at \ieec. Temperature measurements have been made in rather
demanding conditions of environmental thermal stability ---actually,
significantly better than those in the \lca during flight--- which ensure
performance is cleanly assessed. The latest results reported in
section~\ref{lobo-sec2} show that FM parts comply with the requirement
of 10$^{-5}$\,K/$\sqrt{\rm Hz}$ throughout the measuring bandwidth.
Beyond these results, further investigation of system response at
frequencies below 1\,mHz has shown that both \ltp sensors and front-end
electronics maintain a level of noise of 10$^{-5}$\,K/$\sqrt{\rm Hz}$
down to \lisa's lower end at 0.1\,mHz. This is an extremely encouraging
result, even if further research will be needed for \lisa, since
10$^{-5}$\,K/$\sqrt{\rm Hz}$ is already the current thermal stability
requirement inside \lisa's science module, which means a less noisy
temperature measurement has to be implemented.

Magnetic diagnostics are also in place, but improved data analysis
procedures are needed, and currently under investigation at \ieec.
Looking into \lisa, a more efficient sensing setup is clearly necessary,
with more sensors, and placed closer to the TMs. Recent studies show that
this is possible with AMR magnetometers, and preliminary tests indicate
that promising performance can be obtained down to 0.1\,mHz.

The Flight Model of the RM is currently under construction, after the
EQM has been satisfactorily debugged. It will be submitted to milder dose
proton irradiation tests to ensure before final delivery it works properly.

Summing up, the \lpf Diagnostics Subsystem is fully in place. Current
work is ongoing on its integration in the mission Experiment Master Plan,
where full practical functionality will be implemented.

\ack

Many thanks are due to C\'esar Boatella (now in \textsl{CNES}-Toulouse),
our colleagues at \ifae, Mokhtar Chemeissani and Carles Puigdengoles, at
Imperial College, Henrique Ara\'ujo, Diana Shaul and Peter Wass (currently
in Trento) and the \lpf teams in AEI-Hannover and the University of Trento:
all of them have contributed both ideas and effective work to the \dds
development. Financial support from the Spanish Ministry of Education,
contracts ESP2004-01647 and ESP2007-61712, is gratefully acknowledged.
CFS acknowledges support from the Ramón y Cajal Programme of the
Ministry of Education and Science of Spain and by a Marie Curie
International Reintegration Grant (MIRG-CT-2007-205005/PHY) within
the 7th European Community Framework Programme.

\end{document}